\documentclass[pra,twocolumn]{revtex4}
\usepackage{amssymb}
\usepackage{amsfonts}
\usepackage{amsmath}
\usepackage{graphicx}
\usepackage{color}

\begin{document}
\title{Memory~effects~in~multipartite~systems~coupled~by~non-diagonal~dephasing~mechanisms}
\author{Adri\'{a}n A. Budini}
\affiliation{Consejo Nacional de Investigaciones Cient\'{\i}ficas y T\'{e}cnicas
(CONICET), Centro At\'{o}mico Bariloche, Avenida E. Bustillo Km 9.5, (8400)
Bariloche, Argentina, and Universidad Tecnol\'{o}gica Nacional (UTN-FRBA),
Fanny Newbery 111, (8400) Bariloche, Argentina}
\date{\today }

\begin{abstract}
The developing of (non-Markovian) memory effects strongly depends on the
underlying system-environment dynamics. Here we study this problem in
multipartite arrangements where all subsystems are coupled to each other by
non-diagonal Markovian (Lindblad) dephasing mechanisms. Taking as system and
environment arbitrary sets of complementary subsystems it is shown that both
operational and non-operational approaches to quantum non-Markovianity can
be characterized in an exact analytical way. Similarly to previous studies
about dissipative-entanglement-generation in this kind of dynamics [Seif,
Wang, and Clerk, Phys. Rev. Lett. \textbf{128}, 070402 (2022)], we found
that memory effects can only emerge when a time-reversal symmetry is broken.
Nevertheless, it is also found that departures from Markovianity can
equivalently be represented through a statistical mixture of Markovian
dephasing dynamics, which does not involve any system-environment
entanglement. Specific bipartite and multipartite dynamics exemplify the
main general results.
\end{abstract}

\maketitle

\section{Introduction}

In the last years remarkable advancements has been achieved in the study and
characterization of open quantum systems~\cite{breuerbook,vega,wiseman}. In
particular, the old association of memory effects with time-convoluted
contributions in the time-evolution of the system density matrix~\cite%
{vanKampen} has been surpassed. Instead, \textit{quantum non-Markovianity}
can now be understood from two alternative powerful theoretical main
streamlines.

First, in \textit{non-operational approaches}, memory effects are only
determined by taking into account the (unperturbed) system density
propagator. Markovianity (memoryless regime) is univocally associated to
quantum semigroup structures~\cite{alicki}. Thus, deviations in the
propagator properties with respect to this reference are used to quantify
the magnitude of memory effects~\cite{BreuerReview,plenioReview}. Diverse
witnesses have been proposed, such as the trace distance between two initial
states~\cite{BreuerFirst}, the divisibility of the propagator~\cite%
{divisible}, non-Markovianity degree~\cite{degree}, the quantum regression
theorem~\cite{QRT,QRTOld}, and the sign of the rates in a canonical Lindblad
structure~\cite{hall}, just to name a few. Secondly, \textit{operational
approaches} have been introduced more recently. Here, the system of interest
is subjected to a set of explicit measurement processes. Markovianity is
related to the usual concept in terms of probabilities~\cite{vanKampen}.
Thus, memory effects are characterized from the joint probabilities of the
measurement outcomes~\cite{modi,budiniCPF,bonifacio,han,ban,rio,hefei}.

Both operational and non-operational approaches to quantum non-Markovianity
provide complementary and valid frames to understand memory effects.
Nevertheless, different conclusions can be obtained in some cases. For
example, the conditions under which memory effects can be interpreted in
terms of an environment-to-system backflow of information strongly differ in
both schemes~\cite{megier,maximal,petruccione,amato,goan,EntroBack,backflow}.

In the operational approach the absence of any (physical)
environment-to-system backflow of information was associated to
(non-Markovian) casual bystander environments~\cite{casual}, that is, those
whose self-dynamics do not depend at all on the system degrees of freedom. A
measurement based procedure enables to detecting this condition~\cite{BIF}.
In addition, it allows to determine if the environment action, when
considering the outcome statistics, can be represented in terms of this kind
of \textquotedblleft passive environments,\textquotedblright\ such as for
example \textit{statistical mixtures of different Markovian evolutions}
(unitary~\cite{backflow} or dissipative Lindblad ones). This kind of
evolutions, in the unitary case, has also been studied from the perspective
of memory effects in non-operational approaches~\cite{poland}.
Interestingly, with a totally different motivation, the possibility of
representing an open quantum system dynamics in terms of a statistical
mixture (random noisy ensembles) of Markovian evolutions has been associated
to the \textit{classicality} of the system-environment interaction~\cite%
{nori,Chen,franco,ChenChen,lika,Szanko}.

All previous issues have been mainly discussed in single open quantum
systems. Nevertheless, given that quantum information becomes relevant when
implemented in \textit{multipartite arrangements}, there has been a growing
interest in the study of this kind of dynamics (from an open system
perspective), both from unitary and dissipative (or effective) underlying
descriptions~\cite%
{palma,rioExp,kike,campo,guff,poletti,sabrina,multi,florescu,daley,Clerck}.
The main goal of this work is to contribute to this research line by
providing a full characterization of quantum non-Markovianity, jointly with
the previous topics, in a class of multipartite dissipative dynamics~\cite%
{Clerck}.

In Ref.~\cite{Clerck} the authors study a multipartite qubit dynamics, where
all subsystems are coupled between them by non-diagonal dephasing
mechanisms. Depending on the dimensionality (number of qubits) and coupling
parameters the dynamics may lead to the emergence of transient \textit{%
multipartite entanglement}~\cite{entanglement}. This property is read as a
signature of the nonclassicality of the evolution. Here, by considering both
Hamiltonian and dissipative couplings [see Eqs.~(\ref{TotalLindblad}) and~(%
\ref{Hamiltonian})] we show that, for any kind of subsystems (qubits or
arbitrary ones), the multipartite dynamics can be diagonalized in an exact
analytical way. Consequently, both operational and non-operational
approaches to quantum non-Markovianity can be tackled in the same way.
Similarly to the study of entanglement generation~\cite{Clerck}, we find
that the break of a time-reversal symmetry plays a fundamental role when
considering the emergence of memory effects. In contrast, the possibility of
representing the dynamic of an arbitrary set of subsystems in terms of a
statistical mixture of Markovian dephasing dynamics is also established.

The paper is outlined as follows. In Sec.~II the multipartite dynamics is
solved in an exact way. Introducing an arbitrary system-environment
splitting, conditions for the emergence of memory effects in non-operational
approaches are obtained. In Sec.~III we characterize memory effects when
considering successive measurement processes performed over the subsystems
of interest. In Sec.~IV we study bipartite and multipartite specific
examples. In Sec.~V we provide the Conclusions. Extensions and calculation
details are provided in the Appendixes.

\section{Multipartite non-diagonal dephasing dynamics}

We consider a multipartite system consisting of an arbitrary set of $n$
subsystems. In general, each one has associated a (possibly different)
Hilbert space $\mathcal{H}_{i}.$ Hence, the total Hilbert space is $\mathcal{%
H}=\mathcal{H}_{1}\otimes \mathcal{H}_{2}\cdots \otimes \mathcal{H}_{n}.$ By
assumption, the total density matrix $\rho _{t}$ obeys the evolution%
\begin{equation}
\frac{d\rho _{t}}{dt}=-i[H,\rho _{t}]+\sum_{i,j}\Gamma _{ij}(S^{(i)}\rho
_{t}S^{(j)}-\frac{1}{2}\{S^{(j)}S^{(i)},\rho _{t}\}_{+}).
\label{TotalLindblad}
\end{equation}%
The indexes $i=1,2,\cdots n$ and $j=1,2,\cdots n$ label the subsystems. In
addition, $S^{(i)}$ is an arbitrary Hermitian operator $(S^{(i)}=S^{(i)%
\dagger })$ acting on each subsystem Hilbert space $\mathcal{H}_{i}.$ $%
\{A,B\}_{+}$ denotes an anticommutator operation between two arbitrary
operators $A$ and $B.$ Hence, the second term in Eq.~(\ref{TotalLindblad})
is a Lindblad contribution that introduces a dissipative coupling between
all pairs of subsystems. For guarantying the complete positive nature of the
solution map, the complex (rate) coefficients $\{\Gamma _{ij}\}$ must to
constitute a positive definite Hermitian matrix~\cite{breuerbook}. The total
Hamiltonian is assumed to be%
\begin{equation}
H=\frac{1}{2}\sum_{i,j}h_{ij}S^{(i)}S^{(j)},  \label{Hamiltonian}
\end{equation}%
where $h_{ij}$ are real coefficients. They scale a unitary coupling between
all subsystems. The model studied in Ref.~\cite{Clerck} is recovered by
taking all subsystems as qubits with $S^{(i)}$ the $z$-Pauli matrix in $%
\mathcal{H}_{i}.$

\subsection{Density matrix solution}

An explicit expression for $\rho _{t}$ can be obtained by introducing an
appropriate basis for the full Hilbert space. Given that each operator $%
S^{(i)}$ is Hermitian, its eigenvectors $\{|s_{i}\rangle \}$ provide a
natural basis for $\mathcal{H}_{i},$ where $S^{(i)}|s_{i}\rangle
=s_{i}|s_{i}\rangle .$ The set $\{s_{i}\}$ are the corresponding
eigenvalues. The basis $\{|\mathbf{s}\rangle \}$ of the full multipartite
Hilbert space $\mathcal{H}$ is then taken as%
\begin{equation}
|\mathbf{s}\rangle \equiv |s_{1}\cdots s_{n}\rangle =|s_{1}\rangle \otimes
|s_{2}\rangle \otimes \cdots \otimes |s_{n}\rangle .  \label{EseKet}
\end{equation}%
With the previous definitions, the dephasing nature of Eq.~(\ref%
{TotalLindblad}) can explicitly be shown, that is, the matrix elements of $%
\rho _{t}$ do not couple to each other. In fact, taking two arbitrary basis
states, $|\mathbf{s}\rangle $ and $|\mathbf{\tilde{s}}\rangle ,$ and using
that $S^{(i)}|\mathbf{s}\rangle =s_{i}|\mathbf{s}\rangle ,$ from Eq.~(\ref%
{TotalLindblad}) we get%
\begin{equation}
\frac{d}{dt}\langle \mathbf{\tilde{s}}|\rho _{t}|\mathbf{s}\rangle =-\Phi _{%
\mathbf{\tilde{s}},\mathbf{s}}\langle \mathbf{\tilde{s}}|\rho _{t}|\mathbf{s}%
\rangle .  \label{CoherenceEvolution}
\end{equation}%
The complex coefficients $\Phi _{\mathbf{\tilde{s}},\mathbf{s}}$ are given by%
\begin{equation}
\Phi _{\mathbf{\tilde{s}},\mathbf{s}}=i(\Omega _{\mathbf{\tilde{s}}}-\Omega
_{\mathbf{s}})+\Upsilon _{\mathbf{\tilde{s}},\mathbf{s}}.  \label{PHI}
\end{equation}%
Here, the \textquotedblleft frequencies\textquotedblright\ $\Omega _{\mathbf{%
s}}$ are induced by the Hamiltonian contribution~(\ref{Hamiltonian}), being
defined as%
\begin{equation}
\Omega _{\mathbf{s}}=\frac{1}{2}\sum_{i,j}h_{ij}s_{i}s_{j}.
\label{frequencies}
\end{equation}%
The contribution $\Upsilon _{\mathbf{\tilde{s}},\mathbf{s}},$ induced by the
non-diagonal Lindblad term in Eq.~(\ref{TotalLindblad}), after a simple
algebra, can be written as%
\begin{equation}
\Upsilon _{\mathbf{\tilde{s}},\mathbf{s}}=\sum_{i,j}(\tilde{s}_{i}-s_{i})%
\frac{\Gamma _{ij}}{2}(\tilde{s}_{j}-s_{j})+\sum_{i,j}\frac{\Gamma _{ij}}{2}(%
\tilde{s}_{j}s_{i}-\tilde{s}_{i}s_{j}).  \label{rates}
\end{equation}%
Notice that the first and second sum contributions depend respectively on
the \textit{real and imaginary parts of the coefficients }$\{\Gamma _{ij}\}.$
These properties follow straightforwardly from the index interchange $%
i\leftrightarrow j.$

The matrix element behavior defined by Eq.~(\ref{CoherenceEvolution}) can be
integrated straightforwardly. Consequently, the multipartite state $\rho
_{t} $ can explicitly be written as%
\begin{equation}
\rho _{t}=\sum_{\mathbf{s},\mathbf{\tilde{s}}}|\mathbf{\tilde{s}}\rangle
\langle \mathbf{\tilde{s}}|\rho _{0}|\mathbf{s}\rangle \langle \mathbf{s}%
|\exp [-\Phi _{\mathbf{\tilde{s}},\mathbf{s}}t],  \label{RhoSolution}
\end{equation}%
where $\rho _{0}$ is the initial multipartite state. Notice that populations
do not evolve in time, $\langle \mathbf{s}|\rho _{t}|\mathbf{s}\rangle =$ $%
\langle \mathbf{s}|\rho _{0}|\mathbf{s}\rangle .$ This property follows from
Eqs.~(\ref{frequencies}) and~(\ref{rates}), which imply $\Phi _{\mathbf{s},%
\mathbf{s}}=0.$ The expression~(\ref{RhoSolution}) allows us to analyze
diverse aspects of the dynamics in an explicit analytical way. It is valid
for arbitrary operators $\{S^{(i)}\}$ and coupling matrixes $\{h_{ij}\}$ and 
$\{\Gamma _{ij}\}.$ Interestingly, an analytical solution can also be found
even when the unitary and dissipative coupling in Eq.~(\ref{TotalLindblad})
are defined by more than two (multipartite) operators (see Appendix~\ref%
{apendice}).

\subsection{System-environment splitting}

In Eq.~(\ref{TotalLindblad}) all subsystems play the same role. In order to
analyze memory effects an arbitrary system-environment splitting must be
introduced. Thus, the total Hilbert space is written as $\mathcal{H}=%
\mathcal{H}_{S}\otimes \mathcal{H}_{B}.$ We consider that $n_{S}$ and $n_{B}$
subsystems, with $n_{S}+n_{B}=n,$ define the system ($\mathcal{H}_{S}$) and
\textquotedblleft bath\textquotedblright\ ($\mathcal{H}_{B}$) Hilbert space
respectively. When $n_{B}>1$ the environment is a multipartite one. In a
similar way, without loss of generality, each element of the basis $\{|%
\mathbf{s}\rangle \}$ [Eq.~(\ref{EseKet})] is rewritten as%
\begin{equation}
|\mathbf{s}\rangle \rightarrow |\mathbf{sb}\rangle \equiv |s_{1}\cdots
s_{n_{S}}\rangle \otimes |b_{1}\cdots b_{n_{B}}\rangle .  \label{spliting}
\end{equation}%
Introducing the change of notation $\rho _{t}\rightarrow \rho _{t}^{se},$
the total density matrix defined by Eq.~(\ref{RhoSolution}) is re-expressed
as%
\begin{equation}
\rho _{t}^{se}=\sum_{\mathbf{s},\mathbf{\tilde{s},b},\mathbf{\tilde{b}}}|%
\mathbf{\tilde{s}\tilde{b}}\rangle \langle \mathbf{\tilde{s}\tilde{b}}|\rho
_{0}^{se}|\mathbf{sb}\rangle \langle \mathbf{sb}|\exp [-\Phi _{\mathbf{%
\tilde{s}\tilde{b}},\mathbf{sb}}t].  \label{RhoSE}
\end{equation}%
Here, $\Phi _{\mathbf{\tilde{s}\tilde{b}},\mathbf{sb}}$ follows from Eq.~(%
\ref{PHI}) after introducing the splitting $\mathbf{\tilde{s}}\rightarrow (%
\mathbf{\tilde{s}},\mathbf{\tilde{b}})$ and $\mathbf{s}\rightarrow (\mathbf{s%
},\mathbf{b}),$ that is,%
\begin{equation}
\Phi _{\mathbf{\tilde{s}\tilde{b}},\mathbf{sb}}=i(\Omega _{\mathbf{\tilde{s}%
\tilde{b}}}-\Omega _{\mathbf{sb}})+\Upsilon _{\mathbf{\tilde{s}\tilde{b}},%
\mathbf{sb}}.  \label{RatesSB}
\end{equation}%
The frequency terms associated to the unitary evolution immediately lead to $%
\Omega _{\mathbf{s}}\rightarrow \Omega _{\mathbf{sb}},$ with%
\begin{equation}
\Omega _{\mathbf{sb}}=\Omega _{\mathbf{s}}+\Omega _{\mathbf{b}}+\sum_{i\in
S,j\in B}\Big{(}\frac{h_{ij}+h_{ji}}{2}\Big{)}s_{i}b_{j}.
\end{equation}%
The sum indexes $i\in S$ and $j\in B$ run over the subsystems associated to
the system and the environment respectively. The \textquotedblleft
non-coupling\textquotedblright\ contributions $\Omega _{\mathbf{s}}$ and $%
\Omega _{\mathbf{b}}$ are given by Eq.~(\ref{frequencies}) but restricting
the sum indexes as $(i,j)\in S$ and $(i,j)\in B$ respectively. On the other
hand, the contribution $\Upsilon _{\mathbf{\tilde{s}},\mathbf{s}}\rightarrow
\Upsilon _{\mathbf{\tilde{s}\tilde{b}},\mathbf{sb}}$ can be written as%
\begin{equation}
\Upsilon _{\mathbf{\tilde{s}\tilde{b}},\mathbf{sb}}=\Upsilon _{\mathbf{%
\tilde{s}},\mathbf{s}}+\Upsilon _{\mathbf{\tilde{b}},\mathbf{b}}+\chi _{%
\mathbf{\tilde{s}\tilde{b}},\mathbf{sb}}.  \label{GamaAbiertaSB}
\end{equation}%
The terms $\Upsilon _{\mathbf{\tilde{s}},\mathbf{s}}$ and $\Upsilon _{%
\mathbf{\tilde{b}},\mathbf{b}}$ have the same structure than Eq.~(\ref{rates}%
) with the restrictions $(i,j)\in S$ and $(i,j)\in B$ respectively. The
contribution $\chi _{\mathbf{\tilde{s}\tilde{b}},\mathbf{sb}}$ introduces
the system-environment coupling. It reads%
\begin{eqnarray}
\chi _{\mathbf{\tilde{s}\tilde{b}},\mathbf{sb}} &=&\sum_{i\in S,j\in B}(%
\tilde{s}_{i}-s_{i})\Big{(}\frac{\Gamma _{ij}+\Gamma _{ji}}{2}\Big{)}(\tilde{%
b}_{j}-b_{j})  \notag \\
&&+\sum_{i\in S,j\in B}\Big{(}\frac{\Gamma _{ij}-\Gamma _{ji}}{2}\Big{)}(%
\tilde{b}_{j}s_{i}-\tilde{s}_{i}b_{j}).  \label{InteractionTerm}
\end{eqnarray}%
Notice that the sum terms depend respectively on the real and imaginary
parts of the matrix $\{\Gamma _{ij}\}.$

\subsection{System dynamics}

Of special interest is to determine the system density matrix, which is
obtained by tracing out the environment degrees of freedom, $\rho
_{t}^{(s)}\equiv \mathrm{Tr}_{e}[\rho _{t}^{se}].$ Similarly, for the
environment $\rho _{t}^{(e)}\equiv \mathrm{Tr}_{s}[\rho _{t}^{se}].$ By
taking separable initial conditions $\rho _{0}^{se}=\rho _{0}^{(s)}\otimes
\rho _{0}^{(e)},$ from Eq.~(\ref{RhoSE}) we get%
\begin{equation}
\rho _{t}^{(s)}=\sum_{\mathbf{s},\mathbf{\tilde{s}}}f_{\mathbf{\tilde{s}s}%
}(t)|\mathbf{\tilde{s}}\rangle \langle \mathbf{\tilde{s}}|\rho _{0}^{(s)}|%
\mathbf{s}\rangle \langle \mathbf{s}|,  \label{RhoSyst}
\end{equation}%
where the set of functions $\{f_{\mathbf{\tilde{s}s}}(t)\}$ is given by%
\begin{equation}
f_{\mathbf{\tilde{s}s}}(t)=\sum_{\mathbf{b}}\langle \mathbf{b}|\rho
_{0}^{(e)}|\mathbf{b}\rangle \exp (-t\Phi _{\mathbf{\tilde{s}b},\mathbf{sb}%
}).  \label{EfeS}
\end{equation}%
From these expressions it is simple to realize that a dephasing mechanism
also characterizes the system dynamics, where the decay of the system
coherences $\langle \mathbf{\tilde{s}}|\rho _{t}^{(s)}|\mathbf{s}\rangle $
is defined by the functions $f_{\mathbf{\tilde{s}s}}(t).$ Consistently,
given that $f_{\mathbf{ss}}(t)=\sum_{\mathbf{b}}\langle \mathbf{b}|\rho
_{0}^{(e)}|\mathbf{b}\rangle =1,$ the populations do not change with time, $%
\langle \mathbf{s}|\rho _{t}^{(s)}|\mathbf{s}\rangle =\langle \mathbf{s}%
|\rho _{0}^{(s)}|\mathbf{s}\rangle .$ In Appendix~\ref{BathDynamics} we
explicitly write the environment state.

In contrast to Eq.~(\ref{RhoSolution}), the coherences behavior defined by $%
f_{\mathbf{\tilde{s}s}}(t)$ strongly depart from an (complex) exponential
one. This property anticipates the presence of memory effects, which is
supported by characterizing the time-evolution of $\rho _{t}^{(s)}.$ The
most general time-dependent (dephasing) evolution consistent with Eq.~(\ref%
{RhoSyst}) can be written as%
\begin{equation}
\frac{d\rho _{t}^{(s)}}{dt}=\mathcal{L}_{t}[\rho _{t}^{(s)}]+\sum_{\mathbf{%
\tilde{s}s}}\gamma _{t}^{\mathbf{\tilde{s}s}}(\Pi _{\mathbf{\tilde{s}}}\rho
_{t}^{(s)}\Pi _{\mathbf{s}}-\frac{1}{2}\{\Pi _{\mathbf{s}}\Pi _{\mathbf{%
\tilde{s}}},\rho _{t}^{(s)}\}_{+}),  \label{SLindbladTime}
\end{equation}%
where we have introduced the system projectors $\Pi _{\mathbf{s}}\equiv |%
\mathbf{s}\rangle \langle \mathbf{s}|$ and $\mathcal{L}_{t}[\rho
_{t}^{(s)}]\equiv -i[H_{t}^{(s)},\rho _{t}^{(s)}],$ with Hamiltonian%
\begin{equation}
H_{t}^{(s)}=\frac{1}{2}\sum_{\mathbf{s}}\omega _{t}^{\mathbf{s}}|\mathbf{s}%
\rangle \langle \mathbf{s}|.
\end{equation}%
The set of (time-dependent) frequencies $\{\omega _{t}^{\mathbf{s}}\}$ and
the Hermitian matrix of (complex) coefficients $\{\gamma _{t}^{\mathbf{%
\tilde{s}s}}\}$\ can be determined after knowing the set of functions $\{f_{%
\mathbf{\tilde{s}s}}(t)\}$ [Eq.~(\ref{EfeS})]. From Eq.~(\ref{SLindbladTime}%
), they are related by the equations $(\mathbf{\tilde{s}}\neq \mathbf{s})$%
\begin{equation}
\frac{df_{\mathbf{\tilde{s}s}}(t)}{dt}=-\frac{1}{2}[i(\omega _{t}^{\mathbf{%
\tilde{s}}}-\omega _{t}^{\mathbf{s}})+(\gamma _{t}^{\mathbf{\tilde{s}\tilde{s%
}}}+\gamma _{t}^{\mathbf{ss}})-2\gamma _{t}^{\mathbf{\tilde{s}s}}]f_{\mathbf{%
\tilde{s}s}}(t).  \label{GamaDeTime}
\end{equation}%
Therefore, the unknown functions $\{\omega _{t}^{\mathbf{s}}\}$ and $%
\{\gamma _{t}^{\mathbf{\tilde{s}s}}\}$ can be determinated from $\{[1/f_{%
\mathbf{\tilde{s}s}}(t)](d/dt)f_{\mathbf{\tilde{s}s}}(t)\}.$

\subsection{Necessary condition for the development of memory effects}

In \textit{non-operational approaches }to quantum non-Markovianity~\cite%
{BreuerReview,plenioReview}, when the matrix $\{\gamma _{t}^{\mathbf{\tilde{s%
}s}}\}$ in Eq.~(\ref{SLindbladTime}) is positive definite the system
evolution is classified as Markovian. This kind of general characterization
of the matrix $\{\gamma _{t}^{\mathbf{\tilde{s}s}}\}$ cannot be established
in our case of study. Nevertheless, after providing a specific underlying
model [Eq.~(\ref{TotalLindblad})], it can always be calculated in an exact
analytical way.

In spite of the previous limitation, it is possible to establish a \textit{%
necessary condition} for the developing of memory effects. It terms of the
partial diagonal $(\mathbf{\tilde{b}=b})$ multipartite dephasing rates it
reads%
\begin{equation}
\Phi _{\mathbf{\tilde{s}b},\mathbf{sb}}\neq \Phi _{\mathbf{\tilde{s}},%
\mathbf{s}}.  \label{MemoryCondition}
\end{equation}%
In fact, when this condition is not met $[\Phi _{\mathbf{\tilde{s}b},\mathbf{%
sb}}=\Phi _{\mathbf{\tilde{s}},\mathbf{s}}]$ the system coherences behavior
Eq.~(\ref{EfeS}), using that $\sum_{\mathbf{b}}\langle \mathbf{b}|\rho
_{0}^{(e)}|\mathbf{b}\rangle =1,$\ becomes (complex) exponential.
Consequently the system density matrix [Eq.~(\ref{SLindbladTime})] obey a
time-independent \textquotedblleft Markovian\textquotedblright\ Lindblad
equation. We remark that in non-operational approaches the condition~(\ref%
{MemoryCondition}) is necessary but in general not sufficient for the
developing of memory effects.

From the explicit expression for $\Phi _{\mathbf{\tilde{s}\tilde{b}},\mathbf{%
sb}}$ [Eq. (\ref{GamaAbiertaSB})], taking $\mathbf{\tilde{b}=b,}$
straightforwardly it follows%
\begin{eqnarray}
\Phi _{\mathbf{\tilde{s}b},\mathbf{sb}} &=&\Phi _{\mathbf{\tilde{s}},\mathbf{%
s}}+\sum_{i\in S,j\in B}i\Big{(}\frac{h_{ij}+h_{ji}}{2}\Big{)}(\tilde{s}%
_{i}-s_{i})b_{j}  \label{GamaAbiertaExplicita} \\
&&\ \ \ \ \ \ -\sum_{i\in S,j\in B}\Big{(}\frac{\Gamma _{ij}-\Gamma _{ji}}{2}%
\Big{)}(\tilde{s}_{i}-s_{i})b_{j}.  \notag
\end{eqnarray}%
The first contribution has the same structure as Eq.~(\ref{PHI}), $\Phi _{%
\mathbf{\tilde{s}},\mathbf{s}}=i(\Omega _{\mathbf{\tilde{s}}}-\Omega _{%
\mathbf{s}})+\Upsilon _{\mathbf{\tilde{s}},\mathbf{s}},$ but here it only
involves system degrees of freedom. The two remaining sum contributions lead
to memory effects\ [Eq.~(\ref{MemoryCondition})].

In Eq.~(\ref{GamaAbiertaExplicita}), the sum contribution proportional to $%
(h_{ij}+h_{ji})/2$ corresponds to the system-environment coupling induced by
the Hamiltonian term. On the other hand, the dissipative coupling induced by
the non-diagonal structure is proportional to the imaginary part $(\Gamma
_{ij}-\Gamma _{ji})/2$\ of the coupling rates. It is completely independent
of the corresponding real part $(\Gamma _{ij}+\Gamma _{ji}).$ Thus,\textit{\
system-environment correlations induced by the real part of }$\{\Gamma
_{ij}\}$\textit{\ does not lead to memory effects. }In addition\textit{,
memory effects can only emerge when the a time-reversal symmetry is broken. }%
In fact, this symmetry is broken when the matrix $\{\Gamma _{ij}\}$ is a
complex one. Interestingly, the same conditions were found in Ref.~\cite%
{Clerck} when considering the production of transient entanglement.

An relevant conclusion can also be obtained from Eq.~(\ref%
{GamaAbiertaExplicita}). \textit{While the unitary and dissipative couplings
may lead to different system-environment correlations, they may induce
exactly the same non-Markovian system dynamics.} In fact, in Eq.~(\ref%
{GamaAbiertaExplicita}) the dependence of the sum contributions with respect
to the eigenvalues $\{(\tilde{s}_{i}-s_{i})b_{j}\}$ is exactly the same.
Consequently, under the mapping $i(h_{ij}+h_{ji})/2\leftrightarrow -(\Gamma
_{ij}-\Gamma _{ji})/2,$ exactly the same system memory effects are induced
by the unitary and dissipative couplings respectively [see Eqs.~(\ref%
{RhoSyst}) and~(\ref{EfeS})].

\section{Operational approach to quantum non-Markovianity}

In operational approaches to quantum non-Markovianity the\textit{\ system of
interest} is subjected to a set of measurement processes~\cite%
{modi,budiniCPF}. The classification of the dynamics relies on determining
if the corresponding outcome joint-probability fulfills or does not fulfill
a standard Markov definition~\cite{vanKampen}. Interestingly, a full
characterization of this approach can be formulated for the dynamics under
study.

We assume that the system [defined by the splitting~(\ref{spliting})] is
subjected to three successive measurement processes. The goal is to
calculate the joint probability $P(z,y,x)$ where the sets $\{x\},$ $\{y\},$
and $\{z\}$ correspond to the outcomes of each measurement, which are
performed at times $0,$ $t,$ and $t+\tau $ respectively. The measurement
operators are defined as $\{\Pi _{m}\}$ with $m=x,y,z.$ They fulfill the
normalization condition $\sum_{m}\Pi _{m}^{\dagger }\Pi _{m}=\mathrm{I}_{s},$
where $\mathrm{I}_{s}$ is the identity operator in the system Hilbert space.
The intermediate measurement is assumed to be a projective one. In all
cases, the measurements induce the transformation $\rho \rightarrow \rho
_{m},$ where the post measurement states are $\rho _{m}=\Pi _{m}\rho \Pi
_{m}^{\dagger }/\mathrm{Tr}[\Pi _{m}^{\dagger }\Pi _{m}\rho ],$ each case
occurring with probability $P(m)=\mathrm{Tr}[E_{m}\rho ].$ For simplifying
the expressions we denote $E_{m}\equiv \Pi _{m}^{\dagger }\Pi _{m},$ where $%
m=x,y,z.$

\subsection{Joint probability of measurement outcomes}

We maintain the system-environment splitting defined by Eq.~(\ref{spliting}%
). Thus, the corresponding propagator is set by Eq.~(\ref{RhoSE}).
Furthermore, separable initial conditions are assumed $\rho _{0}^{se}=\rho
_{0}^{s}\otimes \rho _{0}^{e}.$ Consequently, the outcome probability for
the first measurement is $P(x)=\mathrm{Tr}_{s}[E_{x}\rho _{0}^{s}],$ while
the post-measurement state is%
\begin{equation}
\rho _{0}^{se}\rightarrow \rho _{x}^{se}=\rho _{x}\otimes \rho _{0}^{e}.
\label{BiparX}
\end{equation}%
Afterwards, during a time interval of duration $t,$ the arrange follows the
dynamics~(\ref{RhoSE}), which induces the transformation $\rho
_{x}^{se}\rightarrow \rho _{x}^{se}(t),$ where%
\begin{equation}
\rho _{x}^{se}(t)=\sum_{\mathbf{s},\mathbf{\tilde{s},b,\tilde{b}}}|\mathbf{%
\tilde{s}\tilde{b}}\rangle \langle \mathbf{\tilde{s}\tilde{b}}|\rho
_{x}\otimes \rho _{0}^{e}|\mathbf{sb}\rangle \langle \mathbf{sb}|\exp
[-t\Phi _{\mathbf{\tilde{s}\tilde{b}},\mathbf{sb}}].  \label{RhoSB_X}
\end{equation}

The conditional probability for the second measurement outcomes $\{y\},$
given that the first measurement outcome is $x,$ is given by $P(y|x)=\mathrm{%
Tr}_{se}[E_{y}\rho _{x}^{se}(t)].$ It can explicitly be written as%
\begin{equation}
P(y|x)=\sum_{\mathbf{s},\mathbf{\tilde{s},b}}\langle \mathbf{s}|E_{y}|%
\mathbf{\tilde{s}}\rangle \langle \mathbf{\tilde{s}}|\rho _{x}|\mathbf{s}%
\rangle \langle \mathbf{b}|\rho _{0}^{e}|\mathbf{b}\rangle \exp [-t\Phi _{%
\mathbf{\tilde{s}b},\mathbf{sb}}].  \label{Py|x}
\end{equation}%
Using that the second measurement is a projective one, the corresponding
post-measurement state is%
\begin{equation}
\rho _{x}^{se}(t)\rightarrow \rho _{yx}^{se}(t)=\rho _{y}\otimes \rho
_{yx}^{e}(t),  \label{RhoBiparYX}
\end{equation}%
where the environment state is%
\begin{eqnarray}
\rho _{yx}^{e}(t) &=&\frac{1}{P(y|x)}\sum_{\mathbf{b,\tilde{b}}}|\mathbf{%
\tilde{b}}\rangle \langle \mathbf{\tilde{b}}|\rho _{0}^{e}|\mathbf{b}\rangle
\langle \mathbf{b}|  \label{RBathConditionalYX} \\
&&\times \sum_{\mathbf{s},\mathbf{\tilde{s}}}\langle \mathbf{s}|E_{y}|%
\mathbf{\tilde{s}}\rangle \langle \mathbf{\tilde{s}}|\rho _{x}|\mathbf{s}%
\rangle \exp [-t\Phi _{\mathbf{\tilde{s}\tilde{b}},\mathbf{sb}}].  \notag
\end{eqnarray}

Finally, the arrange evolves during a time interval $\tau ,$ inducing the
transformation $\rho _{yx}^{se}(t)\rightarrow \rho _{yx}^{se}(t+\tau ).$
Using the propagator defined by Eq.~(\ref{RhoSE}) it follows%
\begin{equation}
\rho _{yx}^{se}(t+\tau )=\sum_{\mathbf{s},\mathbf{\tilde{s},b,\tilde{b}}}|%
\mathbf{\tilde{s}\tilde{b}}\rangle \langle \mathbf{\tilde{s}\tilde{b}}|\rho
_{yx}^{se}(t)|\mathbf{sb}\rangle \langle \mathbf{sb}|\exp [-\tau \Phi _{%
\mathbf{\tilde{s}\tilde{b}},\mathbf{sb}}].
\end{equation}%
The conditional probability for the last measurement outcomes $\{z\},$ given
that the previous ones were $y$ and $x,$ is given by $P(z|y,x)=\mathrm{Tr}%
_{se}[E_{z}\rho _{yx}^{se}(t+\tau )],$ which yields%
\begin{equation}
P(z|y,x)\!=\!\!\sum_{\mathbf{s},\mathbf{\tilde{s},b}}\!\langle \mathbf{s}%
|E_{z}|\mathbf{\tilde{s}}\rangle \!\langle \mathbf{\tilde{s}}|\rho _{y}|%
\mathbf{s}\rangle \!\langle \mathbf{b}|\rho _{yx}^{e}(t)|\mathbf{b}\rangle
\!\exp [-\tau \Phi _{\mathbf{\tilde{s}b},\mathbf{sb}}],  \label{Pz|yx}
\end{equation}%
where $\rho _{yx}^{e}(t)$ is defined by Eq.~(\ref{RBathConditionalYX}).

The previous calculations steps allows to obtain the joint probability for
the set of three measurement outcomes. In fact, from Bayes rule it follows
that $P(z,y,x)=P(z|y,x)P(y|x)P(x).$ Using Eqs.~(\ref{Pz|yx}) and~(\ref{Py|x}%
) we get%
\begin{eqnarray}
P(z,y,x) &=&\sum_{\mathbf{b}}\Big{\{}\Big{[}\sum_{\mathbf{s,\tilde{s}}%
}\langle \mathbf{s}|E_{z}|\mathbf{\tilde{s}}\rangle \langle \mathbf{\tilde{s}%
}|\rho _{y}|\mathbf{s}\rangle \exp (-\tau \Phi _{\mathbf{\tilde{s}b},\mathbf{%
sb}})\Big{]}  \notag \\
&&\ \ \ \ \ \times \Big{[}\sum_{s,\tilde{s}}\langle \mathbf{s}|E_{y}|\mathbf{%
\tilde{s}}\rangle \langle \mathbf{\tilde{s}}|\rho _{x}|\mathbf{s}\rangle
\exp (-t\Phi _{\mathbf{\tilde{s}b},\mathbf{sb}})\Big{]}  \notag \\
&&\ \ \ \ \ \times \langle \mathbf{b}|\rho _{0}^{e}|\mathbf{b}\rangle %
\Big{\}}P(x).  \label{PconjuntaZYX}
\end{eqnarray}%
This final result provides an explicit expression for $P(z,y,x).$ It only
depends on the chosen measurement processes, the initial conditions, and the
characteristic dephasing rates $\Phi _{\mathbf{\tilde{s}b},\mathbf{sb}}$
[defined by Eq.~(\ref{GamaAbiertaExplicita})].

\subsection{Markovian case}

In the operational approach, the dynamics is memoryless if the outcomes
joint probability fulfill the Markov property: $P(z,y,x)=P(z|y)P(y|x)P(x).$
This equality must be valid for \textit{arbitrary measurement processes}. In
general, the expression~(\ref{PconjuntaZYX}) does not fulfill this
condition, which implies a non-Markovian system dynamics. On the other hand,
it is simple to realize that under the condition%
\begin{equation}
\Phi _{\mathbf{\tilde{s}b},\mathbf{sb}}=\Phi _{\mathbf{\tilde{s}},\mathbf{s}%
},  \label{MarkovCondition}
\end{equation}%
the (operational) Markov property is fulfilled for any election of the
measurement processes. In fact, using that $\sum_{\mathbf{b}}\langle \mathbf{%
b}|\rho _{0}^{e}|\mathbf{b}\rangle =1,$ from Eq.~(\ref{PconjuntaZYX}) we get%
\begin{eqnarray}
P(z,y,x) &=&\Big{[}\sum_{\mathbf{s},\mathbf{\tilde{s}}}\langle \mathbf{s}%
|E_{z}|\mathbf{\tilde{s}}\rangle \langle \mathbf{\tilde{s}}|\rho _{y}|%
\mathbf{s}\rangle \exp (-\tau \Phi _{\mathbf{\tilde{s}},\mathbf{s}})\Big{]}
\label{MarkovPJoint} \\
&&\!\!\!\!\!\times \Big{[}\sum_{\mathbf{s},\mathbf{\tilde{s}}}\langle 
\mathbf{s}|E_{y}|\mathbf{\tilde{s}}\rangle \langle \mathbf{\tilde{s}}|\rho
_{x}|\mathbf{s}\rangle \exp (-t\Phi _{\mathbf{\tilde{s}},\mathbf{s}})\Big{]}%
P(x).  \notag
\end{eqnarray}%
The first and second sum contributions can be read as $P(z|y)$ and $P(y|x)$
respectively, which implies the validity of the Markov property $%
P(z,y,x)=P(z|y)P(y|x)P(x).$

We remark that in the operational approach, Eq.~(\ref{MarkovCondition}) is a%
\textit{\ necessary and sufficient condition for Markovianity}. That is, in
contrast to the non-operational approach, here the inequality $\Phi _{%
\mathbf{\tilde{s}b},\mathbf{sb}}\neq \Phi _{\mathbf{\tilde{s}},\mathbf{s}}$
[Eq.~(\ref{MemoryCondition})] guarantees the presence of memory effects.
Taking into account Eq.~(\ref{GamaAbiertaExplicita}), \textit{a
non-vanishing Hamiltonian term }$(h_{ij}+h_{ji})/2$\textit{\ or any
non-vanishing dissipative imaginary coupling }$(\Gamma _{ij}-\Gamma _{ji})/2$%
\textit{\ guarantee the presence of detectable memory effects}. On the other
hand, it is possible\ that condition~(\ref{MarkovCondition}) is fulfilled
but $\Phi _{\mathbf{\tilde{s}\tilde{b}},\mathbf{sb}}\neq \Phi _{\mathbf{%
\tilde{s}},\mathbf{s}}$ $(\mathbf{\tilde{b}}\neq \mathbf{b}).$ As before,
this case emerges when the non-diagonal rate coefficients are real. In fact,
system-environment correlations induced by the real part $(\Gamma
_{ij}+\Gamma _{ji})/2$ do not lead to departure from (operational or
non-operational) Markovianity.

\subsection{Statistical mixture representation}

During the dynamics, system and environment are intrinsically coupled by
their mutual interaction and transient quantum entanglement can be produced~%
\cite{Clerck}. Consistently, the environment state and dynamics depend on
the system degrees of freedom (see Appendix~\ref{BathDynamics}). In
particular, between the successive measurement processes the environment
state is actively modified.

In spite of the previous properties, we notice that the same outcome
probability [Eq.~(\ref{PconjuntaZYX})] can be obtained from an alternative
underlying dynamics. In fact, the expression for $P(z,y,x)$ can be read as a%
\textit{\ statistical mixture (random superposition) of different system
dephasing Markovian dynamics} [compare with Eq.~(\ref{MarkovPJoint})], each
one with dephasing rates $\Phi _{\mathbf{\tilde{s}b},\mathbf{sb}},$ where
the statistical weight of each one is given by the population $\langle 
\mathbf{b}|\rho _{0}^{e}|\mathbf{b}\rangle .$ Thus, one can obtain the same
joint statistics by considering an \textquotedblleft
environment\textquotedblright\ whose participation in the developing of
memory effects is completely passive, which in turn does not involve any
system-environment entanglement. The same affirmation is valid for the
system state $\rho _{t}^{(s)}$ [see Eqs.~(\ref{RhoSyst}) and~(\ref{EfeS})].

The reading of the system dynamics in terms of a statistical mixture of
Markovian dynamics can be seen as a non-unitary extension of the Hamiltonian
ensemble introduced in Ref.~\cite{nori}. Interestingly, this kind of
equivalence can be detected through the measurement scheme. Considering the
results of Ref.~\cite{BIF}, a \textit{random selection} of the system state
after the second measurement should render the statistics Markovian.
Explicitly, in Eq.~(\ref{PconjuntaZYX}), the following two changes are
introduced%
\begin{equation}
\rho _{y}\rightarrow \rho _{\breve{y}},\ \ \ \ \ \ \ \ \ \ E_{y}\rightarrow
\sum_{y}E_{y}=\mathrm{I}_{s}.  \label{RChanges}
\end{equation}%
The first change implies that after the second measurement, the
post-measurement state is randomly chosen $y\rightarrow \breve{y}$ over the
set $\{\rho _{y}\}.$ This change (performed for example with an unitary
transformation) is chosen with an arbitrary conditional probability $\wp (%
\breve{y}|x).$ As a consequence, the original $y$-outcome is disregarded,
property that lead to the corresponding addition $\sum_{y}.$ Introducing in
successive order the changes~(\ref{RChanges}) into Eq.~(\ref{PconjuntaZYX})
it follows%
\begin{eqnarray}
&&\!\!\!\!\!\!\!P(z,\breve{y},x)\overset{r}{=\!}\Big{[}\!\sum_{\mathbf{b},%
\mathbf{s},\mathbf{\tilde{s}}}\langle \mathbf{s}|E_{z}|\mathbf{\tilde{s}}%
\rangle \langle \mathbf{\tilde{s}}|\rho _{\breve{y}}|\mathbf{s}\rangle \exp
(-\tau \Phi _{\mathbf{\tilde{s}b},\mathbf{sb}})\langle \mathbf{b}|\rho
_{0}^{e}|\mathbf{b}\rangle \!\Big{]}  \notag \\
&&\!\!\ \ \ \ \ \ \ \ \ \ \ \ \ \ \times \wp (\breve{y}|x)P(x),
\end{eqnarray}%
where the symbol $\overset{r}{=}$ implies that this equality is only valid
under the steps~(\ref{RChanges}). As expected, this final expression has the
structure $P(z,\breve{y},x)\overset{r}{=}P(z|\breve{y})\wp (\breve{y}%
|x)P(x), $ that is, independently of the measurement process and chosen
probability $\wp (\breve{y}|x),$ a Markov property is induced. In general,
this Markovian property is not fulfilled. When it applies, it provides an
experimental technique \cite{BIF} for detecting when an environment can be
replaced by a passive, or in general, by a casual bystander one \cite{casual}%
. This feature in turn can be read as the absence of any physical
environment-to-system backflow of information.

\section{Examples}

In this section we apply the previous general theoretical approach to some
specific examples. The properties of memory effects are discussed in detail.

\subsection{Bipartite arrangement}

First we consider a bipartite arrangement [Eq.~(\ref{TotalLindblad}) with $%
n=2].$ Therefore, both the system and the environment consist in one single
system. For clarity, their density matrix evolution is explicitly written as%
\begin{equation}
\frac{d\rho _{t}^{se}}{dt}-i[H,\rho _{t}^{se}]+\mathcal{L}[\rho _{t}^{se}],
\label{BipartitoEvolucion}
\end{equation}%
where the bipartite Hamiltonian is%
\begin{equation}
H=\Omega (S\otimes B),  \label{Hache}
\end{equation}%
while the non-diagonal Lindblad contribution $\mathcal{L}[\rho _{t}^{se}]$
is defined by%
\begin{eqnarray}
\mathcal{L}[\rho _{t}^{se}] &=&+\gamma \Big{(}S\rho _{t}^{se}S-\frac{1}{2}%
\{S^{2},\rho _{t}^{se}\}_{+}\Big{)}  \notag \\
&&+\beta \Big{(}B\rho _{t}^{se}B-\frac{1}{2}\{B^{2},\rho _{t}^{se}\}_{+}%
\Big{)}  \label{ELES} \\
&&+\Big{[}\chi \Big{(}S\rho _{t}^{se}B-\frac{1}{2}\{SB,\rho _{t}^{se}\}_{+}%
\Big{)}+h.c.\Big{]}.  \notag
\end{eqnarray}%
In these expressions $S$ and $B$ are arbitrary Hermitian operators acting in
the system and bath Hilbert spaces respectively. The frequency $\Omega $
measure the strength of the unitary system-environment interaction. On the
other hand, the Hermitian matrix%
\begin{equation}
\{\Gamma _{ij}\}=\left( 
\begin{array}{cc}
\gamma & \chi \\ 
\chi ^{\ast } & \beta%
\end{array}%
\right) ,  \label{Gamma3D}
\end{equation}%
sets the dissipative system-environment interaction. Given that $\{\Gamma
_{ij}\}$ must be a positive definite matrix, it follows the constraints $%
\gamma \geq 0,$ $\beta \geq 0,$ and $\gamma \beta -|\chi |^{2}\geq 0.$

Introducing the eigenvectors and eigenvalues $S|s\rangle =s|s\rangle ,$ $%
B|b\rangle =b|b\rangle ,$ the dephasing rates~(\ref{PHI}) under the
splitting~(\ref{spliting}) can be written as [Eq.~(\ref{RatesSB})]%
\begin{eqnarray}
\Phi _{\tilde{s}\tilde{b},sb} &=&i\Omega (\tilde{s}\tilde{b}-sb)-i\chi _{I}(%
\tilde{s}b-s\tilde{b})+\frac{\gamma }{2}(\tilde{s}-s)^{2}  \notag \\
&&+\frac{\beta }{2}(\tilde{b}-b)^{2}+\chi _{R}(\tilde{s}-s)(\tilde{b}-b),
\end{eqnarray}%
where\ the real and imaginary parts of the non-diagonal coupling rate $\chi $
are denoted as $\chi _{R}=\mathrm{Re}[\chi ]$ and $\chi _{I}=\mathrm{Im}%
[\chi ]$ respectively.

For the emerging of system memory effects we have to consider the (partial)
diagonal contribution $\tilde{b}=b$ [see Eqs.~(\ref{RhoSyst}) and~(\ref{EfeS}%
)], which leads to%
\begin{equation}
\Phi _{\tilde{s}b,sb}=-i(\chi _{I}-\Omega )b(\tilde{s}-s)+\frac{\gamma }{2}(%
\tilde{s}-s)^{2}.  \label{PhiForSystem}
\end{equation}%
Consequently the parameters $\beta $ and $\chi _{R}$ does not participate in
the developing of memory effects. This result is consistent with the general
expression~(\ref{GamaAbiertaExplicita}). On the other hand, from the point
of view of the system dynamics the parameters $\Omega $ and $\chi _{I}$ play
exactly the same role. In fact, the sign of both parameters is arbitrary.
Nevertheless, notice that the underlying coupling processes associated to
these two constants, and the system-environment correlations induced by each
one, are different in general.

\subsubsection*{Two qubits}

As an specific example we consider that both subsystems are qubits. For
simplicity both operators $S$ and $B$ are taken as a $z$-Pauli matrix ($%
\sigma _{z}$) in the corresponding Hilbert spaces. Thus, $s=\pm 1$ and $%
b=\pm 1.$ Using the partial transpose criteria \cite{entanglement} in Eq.~(%
\ref{BipartitoEvolucion}), it follows that system-environment entanglement
can only be induced by the Hamiltonian $H$ [Eq.~(\ref{Hache})].
Complementarily, the dissipative non-diagonal coupling is unable to generate
entanglement in this case.

The system density matrix [Eq.~(\ref{RhoSyst})] reads%
\begin{equation}
\rho _{t}^{(s)}=\left( 
\begin{array}{cc}
p_{+} & c_{0}f(t) \\ 
c_{0}^{\ast }f^{\ast }(t) & p_{-}%
\end{array}%
\right) ,  \label{RhoSsolution}
\end{equation}%
where $p_{\pm }\equiv \langle \pm |\rho _{0}^{(s)}|\pm \rangle $ and $%
c_{0}\equiv \langle +|\rho _{0}^{(s)}|-\rangle $ are respectively the
initial populations and coherence of the system in the eigenbase of $\sigma
_{z}.$ The behavior of the coherences [Eq.~(\ref{EfeS})] is given by%
\begin{equation}
f(t)=e^{-2t\gamma }(q_{+}e^{+it2\underline{\chi }_{I}}+q_{-}e^{-it2%
\underline{\chi }_{I}}),
\end{equation}%
where $q_{\pm }\equiv \langle \pm |\rho _{0}^{(e)}|\pm \rangle $ are the
initial populations of the environment. In this expression and the following
ones, for shortening the expression we introduced the parameter $\underline{%
\chi }_{I}\equiv (\chi _{I}-\Omega ).$

\paragraph{Non-operational approach to memory effects}

From Eq.~(\ref{SLindbladTime}), and consistently with the solution~(\ref%
{RhoSsolution}), the system density matrix time-evolution can be cast in the
form%
\begin{equation}
\frac{d\rho _{t}^{(s)}}{dt}=-i\frac{\omega (t)}{2}[\sigma _{z},\rho
_{t}^{(s)}]+\gamma (t)(\sigma _{z}\rho _{t}^{(s)}\sigma _{z}-\rho
_{t}^{(s)}).  \label{SistemaLindblad}
\end{equation}%
Using the procedure defined by Eq.~(\ref{GamaDeTime}), the time-dependent
frequency is%
\begin{equation}
\omega (t)=-\frac{2\underline{\chi }_{I}(q_{+}-q_{-})}{%
q_{+}^{2}+q_{-}^{2}+2(q_{+}q_{-})\cos (4\underline{\chi }_{I}t)},
\label{Omega}
\end{equation}%
while the time-dependent rate is%
\begin{equation}
\gamma (t)=\gamma +\frac{2\underline{\chi }_{I}(q_{+}q_{-})\sin (4\underline{%
\chi }_{I}t)}{q_{+}^{2}+q_{-}^{2}+2(q_{+}q_{-})\cos (4\underline{\chi }_{I}t)%
}.  \label{Gama}
\end{equation}%
Consistently, both $\omega (t)$ and $\gamma (t)$ only depends on the
characteristic rates $\gamma $ and $\underline{\chi }_{I}.$ On the other
hand, the environment populations $\{q_{\pm }\}$ also govern the emergence
of memory effects. In fact, when $q_{\pm }=1,$ $q_{\mp }=0,$ it follows $%
\omega (t)=\mp 2\underline{\chi }_{I},$ and $\gamma (t)=\gamma .$ Hence, the
system dynamics is Markovian.

In general, the rate $\gamma (t)$ may assume both positive an negative
values, which can be used as a witness of memory effects~\cite{hall}. In
Fig.~1(a) and (b) we plot both the frequency $\omega (t)$\ and the rate $%
\gamma (t)$ for two different values of the (scaled) non-diagonal coupling $%
\underline{\chi }_{I}/\gamma .$ Depending on its value, a transition from
Markovian $[\gamma (t)\geq 0]$\ to non-Markovian dynamics $[\gamma
(t)\gtrless 0]$\ is clearly observed. 
\begin{figure}[tbp]
\includegraphics[bb=56 595 725
1130,angle=0,width=8.5cm]{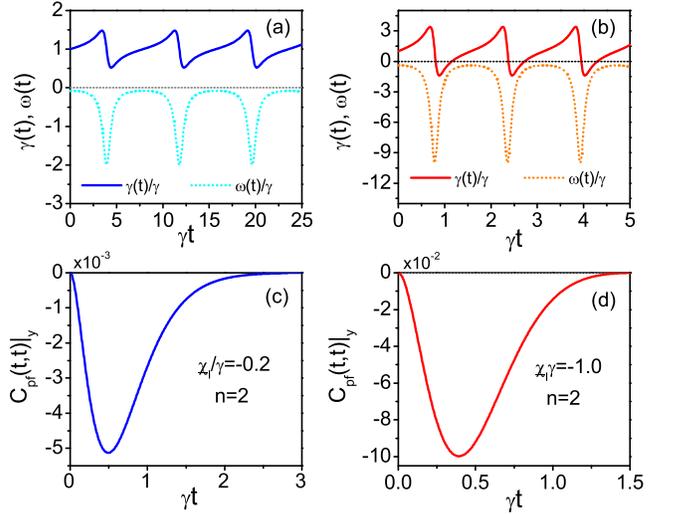}
\caption{Time dependence of the frequency $\protect\omega (t)$ and rate $%
\protect\gamma (t),$ Eqs.~(\protect\ref{Omega}) and~(\protect\ref{Gama})
respectively, jointly with the correlation $C_{pf}(t,\protect\tau )|_{y}$
[Eq.~(\protect\ref{CPFNT_2})] at equal times, for a system coupled to a
single qubit environment, $n=2.$ In (a) and (c) the non-diagonal parameter
is $\protect\underline{\protect\chi }_{I}/\protect\gamma =-0.2,$ while in
(b) and (d) it is $\protect\underline{\protect\chi }_{I}/\protect\gamma %
=-1.0.$ In all cases the environment populations are taken as $q_{+}=0.4,$ $%
q_{-}=0.6,$ while the angle of the intermediate measurement is $\protect\phi %
=\protect\pi /2.$}
\end{figure}

\paragraph{Operational approach to memory effects}

For implementing the operational approach, we assume that the three
measurements are \textit{projective} ones. They are performed successively
in the Bloch directions $\hat{x}-\hat{n}-\hat{x},$ where $\hat{n}=(\cos
(\phi ),\sin (\phi ),0)$ is an arbitrary direction in $\hat{x}$-$\hat{y}$
plane defined by the angle $\phi .$ The successive measurement outcomes are $%
x=\pm 1,$ $y=\pm 1,$ and $z=\pm 1.$ Using the corresponding measurement
projectors $\{\Pi _{m}=|m\rangle \langle m|\}$ associated to each direction~%
\cite{shankar}, from the general expression~(\ref{PconjuntaZYX}), using Eq.~(%
\ref{PhiForSystem}), we get%
\begin{equation}
\frac{P(z,y,x)}{P(x)}=\frac{1}{4}[1+yxf_{\phi }^{(+)}(t)+zyf_{\phi
}^{(-)}(\tau )+zxf_{\phi }(t,\tau )].  \label{ConjuntaExplicita}
\end{equation}%
Here, the auxiliary functions are%
\begin{equation*}
f_{\phi }^{(\pm )}(t)=e^{-2t\gamma }\{q_{+}\cos [2t\underline{\chi }_{I}(\pm
)\phi ]+q_{-}\cos [2t\underline{\chi }_{I}(\mp )\phi ]\},
\end{equation*}%
while the last one is%
\begin{eqnarray*}
\!\!\!\!\!\!\!\!\!f_{\phi }(t,\tau )\!\! &=&\!\!e^{-2\gamma (t+\tau
)}[q_{+}\cos (2t\underline{\chi }_{I}+\phi )\cos (2\tau \underline{\chi }%
_{I}-\phi ) \\
&&\ \ \ \ \ \ \ \ \ \ \!\!\ \!+q_{-}\cos (2t\underline{\chi }_{I}-\phi )\cos
(2\tau \underline{\chi }_{I}+\phi )].
\end{eqnarray*}

Taking $\underline{\chi }_{I}=0$ in Eq.~(\ref{ConjuntaExplicita}), it is
simple to show that $P(z,y,x)$ fulfill a Markov property. A simple way of 
\textit{witnessing} departures of $P(z,y,x)$ from Markovianity is through a
conditional past-future correlation~\cite{budiniCPF}. It is defined as%
\begin{equation}
C_{pf}(t,\tau )|_{y}=\sum_{z,x}zx[P(z,x|y)-P(z|y)P(x|y)].  \label{CPFCorre}
\end{equation}%
Here, $\{z\}$ and $\{x\}$ represent the possible outcomes in the last
(future) and first (past) measurement processes respectively, while the
conditional $y$ is an arbitrary outcome of the intermediate (present)
measurement. From Bayes rule, the Markov property $P(z,y,x)=P(z|y)P(y|x)P(x)$
can be rephrased as a conditional past-future independence, $%
P(z,x|y)=P(z|y)P(x|y),$ which lead to $C_{pf}(t,\tau )|_{y}=0.$ Hence, the
condition $C_{pf}(t,\tau )|_{y}\neq 0$ implies the presence of memory
effects.

Using that $P(z,x|y)=P(z,y,x)/P(y),$ where $P(y)=\sum_{z,x}P(z,y,x),$ from
Eq.~(\ref{ConjuntaExplicita}) the correlation~(\ref{CPFCorre}) reads%
\begin{eqnarray}
C_{pf}(t,\tau )|_{y}\! &=&\!-(4q_{+}q_{-})\sin ^{2}(\phi )e^{-2\gamma
(t+\tau )}  \notag \\
&&\times \sin (2t\underline{\chi }_{I})\sin [2\tau \underline{\chi }_{I}].
\label{CPFNT_2}
\end{eqnarray}%
For simplicity, this result was derived by assuming system initial\
conditions such that $P(x)=1/2.$ In Fig.~1(c) and (d) we plot $C_{pf}(t,\tau
)|_{y}$ at equal measurement time-intervals, $\tau =t.$ The non-diagonal
coupling is in correspondence with Fig.~1(a) and (b) respectively.
Consistently with previous general results, for any non-vanishing value of $%
\underline{\chi }_{I}/\gamma \neq 0,$ in contrast to the negative rate
criteria, here the dynamics is non-Markovian, $C_{pf}(t,\tau )|_{y}\neq 0.$

\subsection{Multipartite environment}

Now we consider a multipartite dynamics [Eq.~(\ref{TotalLindblad}) with $%
n>2].$ As in the previous example all subsystems are taken as qubits. The
\textquotedblleft first qubit\textquotedblright\ is the taken as the system $%
(n_{S}=1)$ and consequently the rest are part of the environment $%
(n_{B}=n-1).$ Similarly, all coupling operators are taken as the $z$-Pauli
matrix ($\sigma _{z}$) in the corresponding Hilbert spaces. The matrix of
rate coefficients $\{\Gamma _{jk}\}$ is taken as%
\begin{equation}
\{\Gamma _{jk}\}=\{(\gamma -\chi )\delta _{jk}\}+\chi |f_{\lambda }\rangle
\langle f_{\lambda }|,  \label{GammaVector}
\end{equation}%
where $\gamma $ and $\chi $ are two real parameters. Furthermore, $\delta
_{jk}$ is the Kronecker delta function. The complex vector is $|f_{\lambda
}\rangle \equiv \sum_{j=1}^{n}e^{2\pi i(j-1)\lambda /n}|\mathbf{e}%
_{j}\rangle ,$ where $\{|\mathbf{e}_{j}\rangle \}_{j=1}^{n}$ is the standard
basis of a vectorial space of dimension $n,$ while $\lambda $ is an
arbitrary dimensionless real parameter. It is simple to check that $\gamma $
and $\chi $ scale the diagonal and non-diagonal elements of $\{\Gamma
_{jk}\} $ respectively. The structure of $\{\Gamma _{jk}\}$ introduced in
Ref.~\cite{Clerck} is recovered when $\chi =\gamma .$

While the developed results allows to characterizing the dynamics in an
exact analytical way, simple expressions are only obtained for special
values of the free parameter $\lambda .$ From now on we take $\lambda =n/4.$
Hence, Eq.~(\ref{GammaVector}) becomes $\Gamma _{jk}=(\gamma -\chi )\delta
_{jk}+\chi \ (i)^{j-1}(-i)^{k-1},$ that is,%
\begin{equation}
\{\Gamma _{jk}\}=\left( 
\begin{array}{ccccc}
\gamma & -i\chi & -\chi & +i\chi & +\chi \\ 
+i\chi & \gamma & -i\chi & -\chi & \ddots \\ 
-\chi & +i\chi & \gamma & -i\chi & \ddots \\ 
-i\chi & -\chi & +i\chi & \gamma & \ddots \\ 
+\chi & \ddots & \ddots & \ddots & \ddots%
\end{array}%
\right) .  \label{GammaUnCuarto}
\end{equation}%
The diagonal elements are equal to $\gamma $\ while the non-diagonal
couplings alternatively change between imaginary $(\pm i\chi )$ and real $%
(\pm \chi )$ values. The positive definite character of $\{\Gamma _{jk}\}$
[Eq.~(\ref{GammaUnCuarto})], which guarantees that the full evolution is a
completely positive one, implies that $\gamma >0$ and the inequalities%
\begin{equation}
-\frac{\gamma }{(n-1)}\leq \chi \leq \gamma ,  \label{Boundary}
\end{equation}%
where$\ n\geq 2.$ In addition, in Eq.~(\ref{TotalLindblad}) we assume $H=0.$

\subsubsection{Entanglement generation}

The generation of entanglement is dephasing dynamics has been characterized
in unitary dynamics~\cite{katarzyna}. For the multipartite non-diagonal
dissipative dynamics [Eq.~(\ref{TotalLindblad})] the corresponding analysis
has been presented previously~\cite{Clerck}. The basic procedure is to
calculate the matrixes $(\tilde{h}_{ij},\tilde{\Gamma}_{ij}),$ which define
the evolution of the total density matrix after transposing the environment
degrees of freedom, $\rho _{t}\rightarrow \rho _{t}^{\intercal }$ (see
Eq.~(4) in Ref.~\cite{Clerck}). Using the partial transpose criterion~\cite%
{entanglement}, it is possible to conclude that when $\{\tilde{\Gamma}%
_{ij}\} $ has negative eigenvalues the dynamics generates transient
entanglement. When $\{\tilde{\Gamma}_{ij}\}$ has positive eigenvalues, the
partial transpose state $\rho _{t}^{\intercal }$ is positive definite and
entanglement generation is not granted~\cite{Clerck,entanglement}.

By determining $\{\tilde{\Gamma}_{ij}\}$ from Eq.~(\ref{GammaUnCuarto}), and
by calculating its eigenvalues for each $n$ (total number of qubits) it is
possible to determinate (numerically) the minimal value of the parameter $%
\chi /\gamma $ that guarantees entanglement generation. In Fig.~2, we plot
the regions where entanglement generation is granted and where
complementarily $\rho _{t}^{\intercal }$ is positive definite. Only for $%
n\geq 3$ there is entanglement generation. Furthermore, positive values of $%
\chi /\gamma $ are necessary, which in turn decreases with $n.$ Both regions
are limited by the constraints defined by Eq.~(\ref{Boundary}). Beyond these
frontiers, the dynamics must be implemented with Hamiltonians contributions. 
\begin{figure}[tbp]
\includegraphics[bb=40 800 485
1132,angle=0,width=7cm]{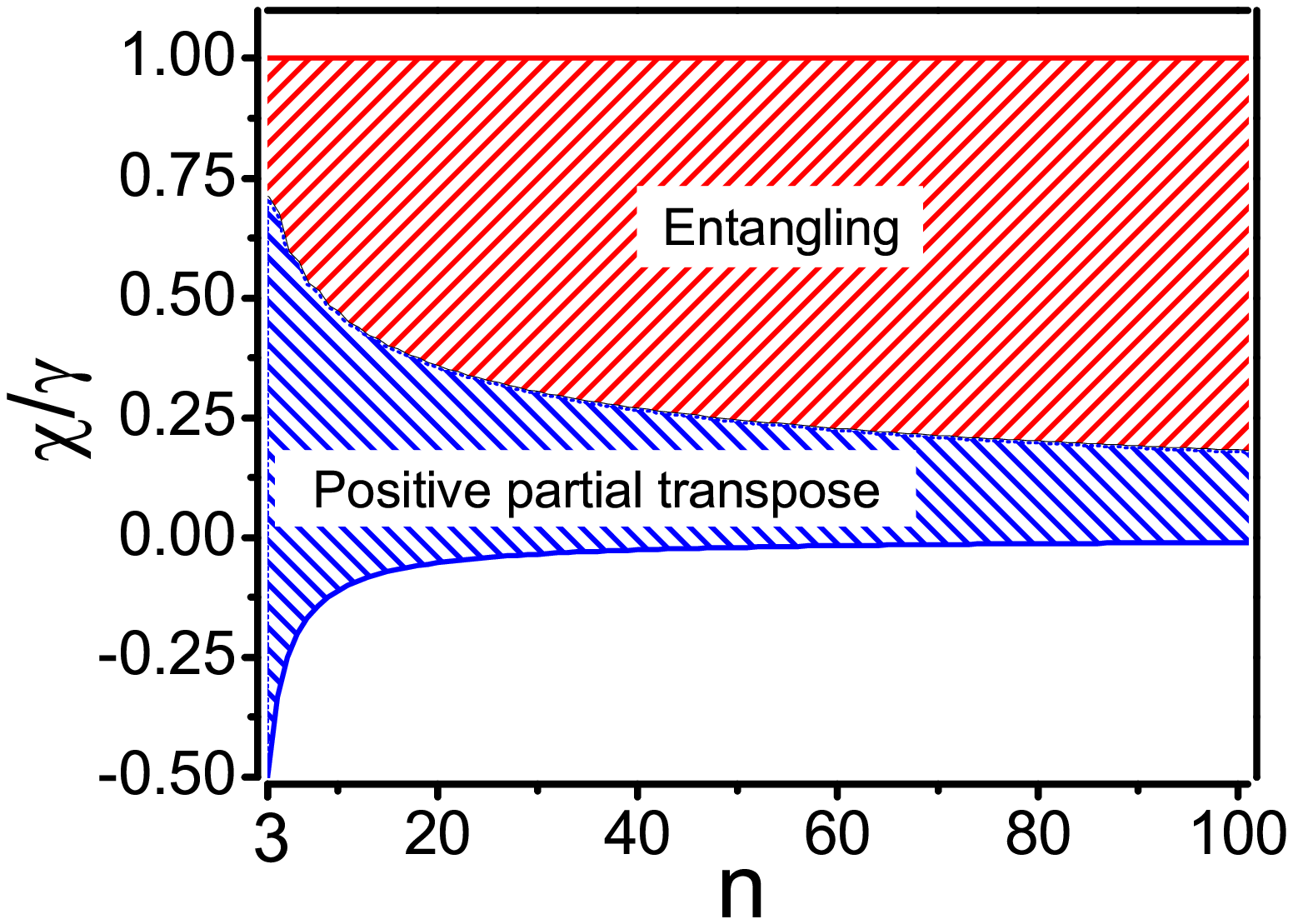}
\caption{Parameter values $\protect\chi /\protect\gamma $ as a function of
the arrangement size $n$ that lead to system-environment entanglement. The
matrix of rate coefficients is given by Eq.~(\protect\ref{GammaUnCuarto}).
The upper and lower boundaries of $\protect\chi /\protect\gamma $ are
defined by Eq.~(\protect\ref{Boundary}). The frontier between the entangling
(negative partial transpose) and the positive partial transpose is determine
numerically (see text).}
\end{figure}

\subsubsection{System memory effects}

Independently of the value of the parameter $\chi /\gamma ,$ the system
state $\rho _{t}^{(s)}$ and its evolution can be written as in Eqs.~(\ref%
{RhoSsolution}) and~(\ref{SistemaLindblad}) respectively. For simplicity, we
assume that all subsystems of the environment begin in an (multipartite)
uncorrelated state, each subsystem having equal upper and lower populations.
Thus, $\langle \mathbf{b}|\rho _{0}^{(e)}|\mathbf{b}\rangle =(1/2)^{n-1}.$
The coherence behavior\ $f(t)$ from Eq.~(\ref{EfeS}), after some algebra, is
given by%
\begin{equation}
f(t)=e^{-2\gamma t}[\cos (2\chi t)]^{\bar{n}},  \label{EfeNqubits}
\end{equation}%
where $\bar{n}\equiv \mathrm{Int}[n/2]$ is the integer part of $(n/2).$ The
dependence on $\bar{n}$ emerges because the non-diagonal elements of $%
\{\Gamma _{jk}\}$ alternate between real and imaginary values [see Eq.~(\ref%
{GammaUnCuarto})]. The time-evolution of $\rho _{t}^{(s)}$ [Eq.~(\ref%
{SistemaLindblad})] is defined with%
\begin{equation}
\omega (t)=0,\ \ \ \ \ \ \ \gamma (t)=\gamma +\bar{n}\chi \tan (2\chi t)].
\label{RateNSpins}
\end{equation}%
The absence of a Hamiltonian contribution $[\omega (t)=0]$ follows from the
equality of the upper and lower populations of each subsystem associated to
the environment. From $\gamma (t),$ we deduce that in \textit{%
non-operational approach} to memory effects, the system dynamics is
non-Markovian whenever $\chi \neq 0.$ Interestingly, this kind of
\textquotedblleft trigonometric eternal non-Markovianity\textquotedblright\
with periodic divergences was also found in a different kind of underlying
multipartite dynamics~\cite{multi}.

For the \textit{operational approach} we choose the same set of measurements
than in the previous bipartite case, $\hat{x}-\hat{n}-\hat{x},$ where the
intermediate one is defined by the angle $\phi .$ The joint probability of
measurement outcomes $P(z,y,x)$ can be written with the structure~(\ref%
{ConjuntaExplicita}). From Eq.~(\ref{PconjuntaZYX}) it follows%
\begin{equation}
f_{\phi }^{(\pm )}(t)=f_{\phi }(t)=f(t)\cos (\phi ),
\end{equation}%
where $f(t)$ is given by Eq.~(\ref{EfeNqubits}), while%
\begin{eqnarray}
f_{\phi }(t,\tau ) &=&\frac{1}{2}e^{-2\gamma (t+\tau )}\{\cos ^{\bar{n}%
}[2\chi (t+\tau )]  \notag \\
&&\quad +\cos (2\phi )\cos ^{\bar{n}}[2\chi (t-\tau )]\}.
\end{eqnarray}%
Consequently, it is simple to check that $P(z,y,x)$ fulfill the Markov
condition only when $\chi =0.$ This property is corroborated by the
conditional past-future correlation [Eq.~(\ref{CPFCorre})], which here can
be written as%
\begin{equation}
C_{pf}(t,\tau )|_{y}=f_{\phi }(t,\tau )-f_{\phi }(t)f_{\phi }(\tau ).
\label{CPF_NT}
\end{equation}%
As before, this result was derived by assuming system initial\ conditions
such that $P(x)=1/2.$

In Fig.~3(a) and (b) we plot the coherence decay~(\ref{EfeNqubits}) and the
conditional past-future correlation~(\ref{CPF_NT}) with $n=6$ and taking
different values of $\chi /\gamma .$ Consistently with their analytical
expressions, the developing of entanglement (see Fig.~2) does not lead to
any significant change in these two objects. This independence follows from
the previous general analysis. In fact, both the system dynamics [Eqs.~(\ref%
{RhoSyst}) and~(\ref{EfeS})] and the outcome statistics [Eq.~(\ref%
{PconjuntaZYX})] can equivalently be obtained from a random superposition of
Markovian dephasing dynamics without involving any multipartite entanglement.

In Fig. 3(c) and (d) we plot the coherence decay~(\ref{EfeNqubits}) and the
conditional past-future correlation~(\ref{CPF_NT}) for different number $n$
of qubits. Given that $\chi /\gamma =1,$ transient entanglement is granted
in all cases. When increasing $n$ the decoherence function $f(t)$ decay in a
faster way and in addition $C_{pf}(t,\tau )|_{y}$ assume higher values,
which can consistently be read as an increasing of system memory effects. 
\begin{figure}[tbp]
\includegraphics[bb=52 590 725
1155,angle=0,width=8.5cm]{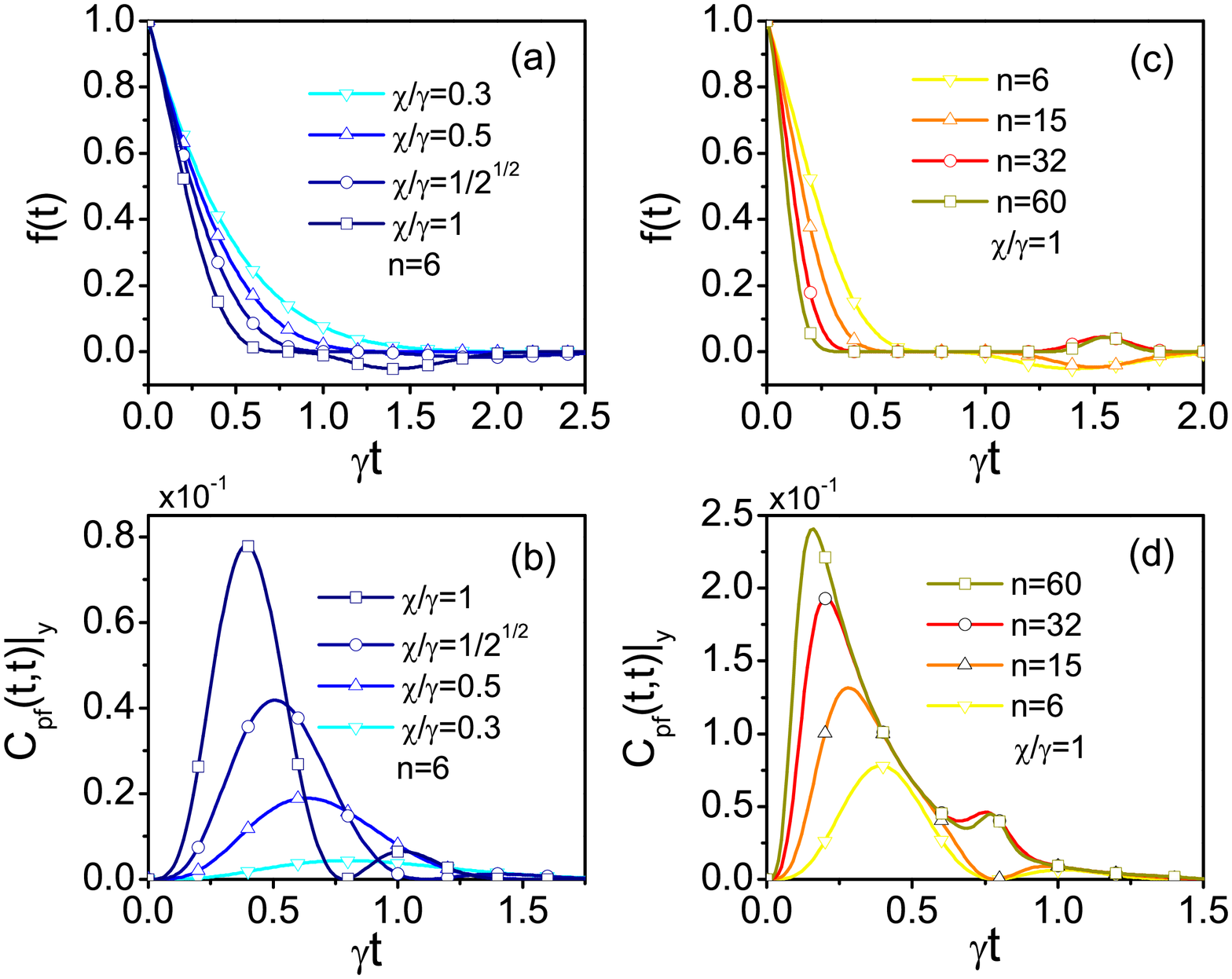}
\caption{Coherence decay $f(t)$ [Eq.~(\protect\ref{EfeNqubits})] and
correlation $C_{pf}(t,\protect\tau )|_{y}$ at equal times [Eq.~(\protect\ref%
{CPF_NT})] for a system coupled to a multipartite environment. The
non-diagonal coupling rate $\protect\chi /\protect\gamma $ and the total
number of qubits $n$ is indicated in each plot. In all cases, the qubits of
the environment begin with equal upper and lower populations. The angle of
the intermediate measurement is $\protect\phi =0.$}
\end{figure}

\subsubsection{Infinite bath size}

The system dynamics can also be characterized in the limit in which the
number of subsystems of the environment become infinite. Nevertheless, for
getting a smooth system coherence decay [Eq.~(\ref{EfeNqubits})], the
non-diagonal dissipative coupling $\chi $ in Eq.~(\ref{GammaUnCuarto}) must
be scaled with the arrangement size $n.$ We assume%
\begin{equation}
\chi \longrightarrow \chi _{n}=g\sqrt{\frac{2}{n}},
\end{equation}%
where $g$ is an arbitrary scaling constant. It is simple to proof that%
\begin{equation}
\lim_{n\rightarrow \infty }[\cos (2\chi _{n}t)]^{\bar{n}}=e^{-2g^{2}t^{2}}.
\end{equation}%
Therefore, for increasing $n,$ the system coherence decay can be fit as%
\begin{equation}
\lim_{n\rightarrow \infty }f(t)=e^{-2\gamma t}\ e^{-2g^{2}t^{2}}.
\end{equation}%
While the diagonal contribution lead to an exponential decay with rate $%
\gamma ,$\ the non-diagonal coupling lead to a Gaussian decay behavior. The
time-dependent rate [Eq.~(\ref{RateNSpins})] becomes $\lim_{n\rightarrow
\infty }\gamma (t)=\gamma +2g^{2}t.$ Remarkably, a similar Gaussian
behaviors can also be obtained from unitary system-environment dynamics~\cite%
{budiniCPF}.

\section{Summary and conclusions}

We studied the emergence and properties of memory effects in a class of
multipartite arrangements where all subsystems are coupled to each other by
non-diagonal Lindblad dephasing mechanisms [Eq.~(\ref{TotalLindblad})]. By
choosing an appropriate basis for the total Hilbert space, the multipartite
density matrix was obtained in an exact analytical way [Eq.~(\ref%
{RhoSolution})]. An arbitrary number of subsystems are associated to the
system of interest, while the rest define its environment. This splitting
[Eq.~(\ref{spliting})] provided the basis for characterizing in an exact way
both non-operational and operational approaches to quantum non-Markovianity.

In non-operational approaches to quantum non-Markovianity, memory effects
are determined from the properties of the system density matrix evolution.
We showed that its general structure can be written as a non-diagonal
time-dependent dephasing evolution [Eq.~(\ref{SLindbladTime})]. Its
characteristic parameters are set by the corresponding system coherence
behaviors [Eq.~(\ref{GamaDeTime})]. A necessary condition for the emergence
of memory effects can be cast in terms of the multipartite dephasing rates
[Eq.~(\ref{MemoryCondition})]. Explicitly, memory effects may be induced by
Hamiltonian couplings or when the dissipative\ coupling breaks a
time-reversal symmetry, that is, the non-diagonal coupling rates must be
complex ones. In these dynamics, these conditions are also necessary for the
development of transient entanglement~\cite{Clerck}.

In operational approaches to quantum non-Markovianity, memory effects are
determine from a set of measurement processes performed over the system of
interest. We calculated in an explicit analytical way the joint probability
of measurement outcomes [Eq.~(\ref{PconjuntaZYX})]. In this case the
previous conditions for the emergence of memory effects become sufficient,
that is, any non-vanishing unitary or dissipative coupling consistent with
the break of time-reversal symmetry lead to departures from Markovianity.

While the multipartite dynamics lead to entanglement generation, we
concluded that this feature is not relevant when considering the properties
of system memory effects. In fact, both the density matrix dynamics and the
statistics of measurement outcomes [Eqs.~(\ref{RhoSyst}) and~(\ref%
{PconjuntaZYX})] can alternatively be obtained from a statistical mixture of
Markovian dephasing evolutions. This equivalent representation does not
involve any entanglement. In addition, in the operational approach, this
property imply that memory effects can be obtained without the occurrence of
any physical environment-to-system backflow of information.

As examples we studied bipartite and multipartite dynamics [with coupling
rates given by Eqs.~(\ref{Gamma3D}) and~(\ref{GammaUnCuarto})], where each
subsystem is a qubit. The properties of the corresponding memory effects
support the previous main results (Figs.~1 to~3).

Understanding the role of system-environment correlations in the developing
of memory effects is a central problem in open quantum system theory. The
present analysis shed light on possible memory features that can emerge in
systems embedded in multipartite dissipative arrangements. Their validity
can in principle be checked in optical setups where this kind of dynamics
can be implemented~\cite{Clerck}.

\section*{Acknowledgments}

The author thanks to Mariano Bonifacio for a critical reading of the
manuscript. This paper was supported by Consejo Nacional de Investigaciones
Cient\'{\i}ficas y T\'{e}cnicas (CONICET), Argentina.

\appendix

\section{Generalization to multipartite couplings\label{apendice}}

In the evolution defined by Eq.~(\ref{TotalLindblad}) the coupling between
the subsystems are bipartite ones, that is, it only involves the action of
two operators: $S^{(i)}$ and $S^{(j)}.$ Multipartite coupling mechanisms can
also be considered, where more than two subsystems are involved. In this
situation, the density matrix can be written as%
\begin{equation}
\frac{d\rho _{t}}{dt}=-i[\text{\b{H}},\rho _{t}]+\sum_{\mathbf{\mu },\mathbf{%
\nu }}\Gamma _{\mathbf{\mu },\mathbf{\nu }}(S_{\mathbf{\mu }}\rho _{t}S_{%
\mathbf{\nu }}-\frac{1}{2}\{S_{\mathbf{\nu }}S_{\mathbf{\mu }},\rho
_{t}\}_{+}),
\end{equation}%
where the indexes are $\mathbf{\mu }=(\mu _{1},\mu _{2},\cdots ,\mu _{n})$
and $\mathbf{\nu }=(\nu _{1},\nu _{2},\cdots ,\nu _{n}).$ The operators $%
\{S_{\mathbf{\mu }}\}$ are defined by the product%
\begin{equation}
S_{\mathbf{\mu }}=S_{\mu _{1}}^{(1)}\otimes \cdots \otimes S_{\mu
_{N}}^{(n)},
\end{equation}%
where each operator $S_{\mu _{i}}^{(i)}$ [defined in $\mathcal{H}_{i}$]
depend on the subindex $\mu _{i}.$ It is defined as%
\begin{equation}
S_{\mu _{i}}^{(i)}\equiv \left\{ 
\begin{array}{c}
S^{(i)}\ \ \ if\ \ \ \mu _{i}=1 \\ 
\mathrm{I}^{(i)}\ \ \ if\ \ \ \mu _{i}=0%
\end{array}%
\right. ,
\end{equation}%
where $\mathrm{I}^{(i)}$\ is the identity operator in the Hilbert space $%
\mathcal{H}_{i}.$ Thus, it is simple to realize that, in contrast to Eq.~(%
\ref{TotalLindblad}), arbitrary multipartite coupling mechanisms are
associated to the coupling rates $\Gamma _{\mathbf{\mu },\mathbf{\nu }}.$
Similarly, the Hamiltonian is taken as \b{H}$=(1/2)\sum_{\mathbf{\mu }}h_{%
\mathbf{\mu }}S_{\mathbf{\mu }},$ where $h_{\mathbf{\mu }}$ are real
coefficients.

In this general situation, by writing $S_{\mu _{i}}^{(i)}=\mu
_{i}S^{(i)}+(1-\mu _{i})\mathrm{I}^{(i)}$ it is simple to check that $S_{%
\mathbf{\mu }}|\mathbf{s}\rangle =\lambda _{\mathbf{s}}^{\mathbf{\mu }}|%
\mathbf{s}\rangle ,$ where the eigenvalue is given by $\lambda _{\mathbf{s}%
}^{\mathbf{\mu }}=\prod_{i=1}^{n}[\mu _{i}s_{i}+(1-\mu _{i})]$ and $|\mathbf{%
s}\rangle $ is defined by Eq.~(\ref{EseKet}). After similar calculations
steps, the density matrix can also be written as in Eq.~(\ref{RhoSolution}).
Here, the frequencies are defined by%
\begin{equation}
\omega _{\mathbf{s}}=\frac{1}{2}\sum_{\mathbf{\mu }}\lambda _{\mathbf{s}}^{%
\mathbf{\mu }}h_{\mathbf{\mu }},
\end{equation}%
while the multipartite dissipative couplings lead to%
\begin{equation}
\Upsilon _{\mathbf{\tilde{s}},\mathbf{s}}=\sum_{\mathbf{\mu },\mathbf{\nu }%
}\Gamma _{\mathbf{\mu },\mathbf{\nu }}(\lambda _{\mathbf{\tilde{s}}}^{%
\mathbf{\mu }}\lambda _{\mathbf{s}}^{\mathbf{\nu }}-\frac{1}{2}\lambda _{%
\mathbf{\tilde{s}}}^{\mathbf{\mu }}\lambda _{\mathbf{\tilde{s}}}^{\mathbf{%
\nu }}-\frac{1}{2}\lambda _{\mathbf{s}}^{\mathbf{\mu }}\lambda _{\mathbf{s}%
}^{\mathbf{\nu }}).
\end{equation}%
By adding and subtracting appropriates terms, this result can be cast with
the same structure than as Eq.~(\ref{rates}).

\section{Environment dynamics\label{BathDynamics}}

A relevant aspect when characterizing memory effects is the environment
dynamics. The system dynamics depends on the environment degrees of freedom
[Eqs.~(\ref{RhoSyst}) and~(\ref{EfeS})]. Given that the system-environment
splitting is arbitrary, a similar property must be valid for the
environment. Specifically, during the dynamics the environment depends on
the system degrees of freedom. In fact, from Eq.~(\ref{RhoSE}) it follows%
\begin{equation}
\rho _{t}^{(e)}=\mathrm{Tr}_{s}[\rho _{t}^{se}]=\sum_{\mathbf{b,\tilde{b}}%
}F_{\mathbf{\tilde{b}b}}(t)|\mathbf{\tilde{b}}\rangle \langle \mathbf{\tilde{%
b}}|\rho _{0}^{(e)}|\mathbf{b}\rangle \langle \mathbf{b}|,  \label{RhoEnvi}
\end{equation}%
where we have introduced the functions%
\begin{equation}
F_{\mathbf{\tilde{b}b}}(t)=\sum_{\mathbf{s}}\langle \mathbf{s}|\rho
_{0}^{(s)}|\mathbf{s}\rangle \exp [-t\Phi _{\mathbf{s\tilde{b}},\mathbf{sb}%
}].
\end{equation}%
Similarly to Eq.~(\ref{EfeS}), here the behavior of the environment
coherences $\{F_{\mathbf{\tilde{b}b}}(t)\}$ is time-dependent and depend on
the system degrees of freedom. Thus, independently of the specific
system-environment splitting [Eq.~(\ref{spliting})] the environment is not a
casual bystander one~\cite{casual}, that is, it dynamically participates in
the generation and developing of system memory effects. Only when\ the
initial environment state $\rho _{0}^{(e)}$ is diagonal in the basis $\{|%
\mathbf{b}\rangle \},$ using that $\Phi _{\mathbf{sb},\mathbf{sb}}=0,$ the
bath dynamics become independent of the system, $\rho _{t}^{(e)}=\rho
_{0}^{(e)}.$

\end{document}